\documentclass[aps,prl,twocolumn,groupedaddress,showpacs]{revtex4}

\usepackage{graphicx, amssymb}
\usepackage{amsmath}
\usepackage{amsfonts}
\usepackage{amssymb}
\usepackage{graphicx}%
\begin{document}
\title{Preparing and probing atomic number states with an atom interferometer}
\author{J. Sebby-Strabley}
\author{B. L. Brown}
\author{M. Anderlini} \thanks{Present address:  INFN, Sezione di Firenze, via Sansone 1, I-50019 Sesto Fiorentino (FI), Italy}
\author{P. J. Lee}
\author{W. D. Phillips}
\author{J. V. Porto} \email{trey@nist.gov}
\affiliation{Joint Quantum Institute, National Institute of
Standards and Technology, and University of Maryland, Gaithersburg,
Maryland 20899, USA }
\author{P. R. Johnson}
\affiliation{Physics Program, American University, Washington D.C.
20016} \affiliation{National Institute of Standards and Technology,
Gaithersburg, Maryland 20899, USA }

\date{\today}

\begin{abstract}
We describe the controlled loading and measurement of
number-squeezed states and Poisson states of atoms in individual
sites of a double well optical lattice. These states are input to an
atom interferometer that is realized by symmetrically splitting
individual lattice sites into double wells, allowing atoms in
individual sites to evolve independently.  The two paths then
interfere, creating a matter-wave double-slit diffraction pattern.
The time evolution of the double-slit diffraction pattern is used to
measure the number statistics of the input state. The flexibility of
our double well lattice provides a means to detect the presence of
empty lattice sites, an important and so far unmeasured factor in
determining the purity of a Mott state.
\end{abstract}
 \maketitle

The optical beam splitter, with its two input and two output modes,
is one of the simplest examples of a two mode quantum system.  At
the quantum level this fundamental system has interesting,
nonclassical behavior such as the quantum interference between
correlated, indistinguishable photons \cite{hom}.  This system
becomes even richer when the particles interact.  While photons may
effectively interact in nonlinear media, the atom optics analog is
naturally interacting. The two mode beam splitter has already been
applied in atom optics experiments; several experiments have split a
trapped Bose-Einstein condensate (BEC) by raising a barrier to
separate the condensate into two independent condensates
\cite{Schumm2005, Shin2005, Shin2004,Albiez2005}. In these
experiments the number of atoms is large, in a regime where
few-particle quantum interference effects cannot be seen. Here we
demonstrate a new few-atom quantum ``beam splitter"
\cite{Mandel2003} and use it to create and analyze classical and
nonclassical states, resulting in interesting few-atom quantum
effects. The ability to create and analyze such states provides a
probe of many-body states in a lattice, a platform for fundamental
studies of few-particle, interacting systems, and is of paramount
importance for quantum computation with neutral atoms.

\begin{figure}[ptb]
\includegraphics[scale=.55]{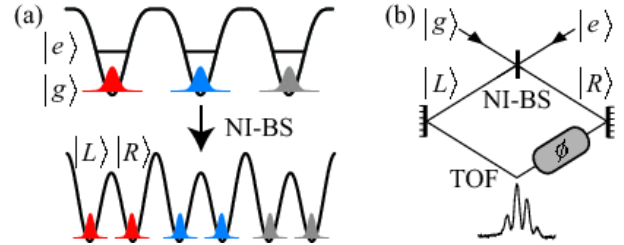}\caption{(color online) (a)  Action of a double well beam
splitter. Different colors indicate the lack of phase coherence
between double well sites.  (b) Optical analog of our atom
interferometer with an example diffraction pattern resulting
from a single input mode.}%
\label{MZ}%
\end{figure}

We realize an atomic analog of the optical two-mode quantum beam
splitter with $^{87}$Rb atoms loaded into an optical lattice of
double wells \cite{SebbyStrabley2006}. This
3D
lattice has a unit cell that can be dynamically transformed between
single well and double well configurations (Fig.~\ref{MZ}a). In
analogy with the optical beam splitter, the input and output modes
in this lattice are either the ground $\left|g\right>$ and first
excited $\left|e\right>$ state of the single well or the ground
state of the left $\left|L\right>$ and right $\left|R\right>$ sites
of the double well. Dynamically switching between these two
configurations effectively creates an array of ``beam splitters"
(BS) coupling these modes (in the limit that higher modes are not
excited). The interactions between atoms and our control of the beam
splitter time scale extends the possible actions beyond those of a
traditional beam splitter.

The speed at which the beam splitter is applied to the atoms
determines whether interactions play a role in the splitting:  if
the BS is applied quickly compared to interaction energies, the
interactions are unimportant. This ``non-interacting" beam splitter
(NI-BS) is analogous to the linear optics case: atoms in a single
input mode (for example, $\left|g\right>$) are divided into two
output modes (here, $\left|L\right>$ and $\left|R\right>$) according
to binomial statistics. In contrast if the BS is applied slowly
compared to interaction energies, the interactions change the
statistics of the distribution between output modes. This
``interacting" beam splitter (I-BS) has no simple optical analog. We
use the NI-BS and I-BS to create and analyze classical Poisson
states and interesting nonclassical few-atom squeezed states. The
ability to controllably and reliably load high fidelity Fock states
is attractive for applications in quantum information, including the
implementation of a two-qubit operation in an optical lattice.

We probe these states using atom interferometry (Fig.~\ref{MZ}). The
NI-BS symmetrically and coherently splits a single lattice site into
a double well. Atoms in individual sites of a double well evolve
independently, analogous to the individual arms of an
interferometer. We interfere the two paths by releasing the atoms
and allowing them to expand and overlap during time-of-flight (TOF)
%
%
The interference from a single pair of sites results in a
double-slit diffraction pattern. In this experiment the initial
states were prepared to have no phase coherence between unit cells;
thus the observed image is an incoherent sum of double-slit
interference patterns from each of the double wells.

The input states were created and analyzed in a double well optical
lattice \cite{SebbyStrabley2006}. We applied this lattice
 to a BEC of $^{87}$Rb atoms in
$5 S_{1/2} \left|F=1,m_F=-1\right>$ created as in \cite{Peil2003}.
The lattice comprises two independent 2D lattices having periodicity
along $\hat{x}$ of either $\lambda$ or $\lambda/2$ (Fig. \ref{MZ}a)
where $\lambda =$ 815 nm  which is red-detuned from the $^{87}$Rb
D-lines. The relative position $\delta x$ of the
``$\lambda$-lattice" and the ``$\lambda/2$-lattice" and their
intensities can be dynamically controlled.  Transforming from the
$\lambda$-lattice to a predominantly $\lambda/2$-lattice with the
minima of the $\lambda/2$-lattice symmetrically straddling the
minima of the $\lambda$-lattice accomplishes the splitting shown in
Fig.~\ref{MZ}.  We use an independent 1D ``vertical lattice" in the
$\hat{z}$ direction to confine atoms to sites of a 3D lattice. The
double well and vertical lattices are focused to $1/e^2$ Gaussian
beam radii of 170 $\mu$m and 250 $\mu$m respectively and have a
relative detuning of 160 MHz. The final depths of the
$\lambda/2$-lattice and vertical lattice are $\approx 30 E_R$ and
$\approx 40 E_R$, respectively ($E_{R}=\hbar ^{2}k^{2}/(2m)=h\times
3.5$ kHz, $k=2\pi/\lambda$, $m$ is the $^{87}$Rb mass). The atomic
momentum distribution in the $x,y$-plane is absorption-imaged after
TOF.

\begin{figure}[ptb]
\includegraphics[scale=.5]{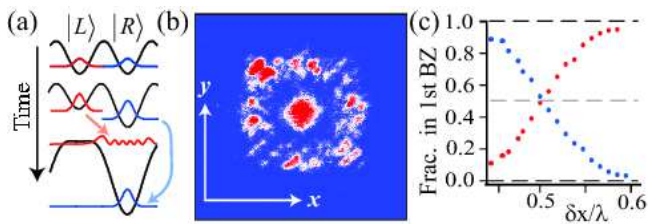}\caption{(color online) (a) Dumping the population
initially in $\left|L\right>$ into highly excited, possibly unbound
bands. (b) The resulting TOF image:  the population initially in
$\left|R\right>$ appears in the first BZ of the $\lambda$-lattice;
the population initially in $\left|L\right>$ appears in higher BZs
as a ring around the first BZ. (c) Calibration of the BS in
Fig.~\ref{MZ}.  After applying the BS at a given value of $\delta
x/\lambda$, the population in the $\left|L\right>$ (red) or
$\left|R\right>$ (blue) sites are measured by dumping the other site
and measuring the fraction of atoms remaining in the first BZ. The
value of $\delta x/\lambda$ where
the two curves cross corresponds to a 50/50 BS.  }%
\label{dumpedBZ}%
\end{figure}

We measure the populations in the two output ports, $\left|L\right>$
and $\left|R\right>$, by selectively imparting energy to atoms in
one of the two sites, which separates the populations in TOF
(Fig.~\ref{dumpedBZ}). Transforming to the $\lambda$-lattice (in
$300\,\mu$s) with its minimum centered over the $\left|R\right>$
site, imparts energy to $\left|L\right>$ atoms; $\left|R\right>$
atoms remain in the ground state of the $\lambda$-lattice
(Fig.~\ref{dumpedBZ}a). The final states' Brillouin zones (BZs),
which are filled, are observed (see for instance
\cite{Kastberg1995,SebbyStrabley2006}) by turning off the lattice in
500\,$\mu$s and imaging after TOF (Fig.~\ref{dumpedBZ}b). The atoms
in the center of the image correspond to the filled first BZ of the
$\lambda$-lattice \cite{SebbyStrabley2006} and were originally in
$\left|R\right>$. The atoms in the ring correspond to higher BZs and
were originally in $\left|L\right>$. This method allows for
\emph{direct measurement} of the atom population in $\left|L\right>$
and $\left|R\right>$. Using this method we can accurately calibrate
the relative position of the two lattices that gives equal splitting
of the NI-BS (shown in Fig.~\ref{dumpedBZ}c). This 50/50 BS was used
for all experiments described here.

To prepare number squeezed states with $N=1$ atom in
$\left|g\right>$ of the $\lambda$-lattice sites, we used a two-stage
``slow" loading procedure. We started with $\approx 2\times 10^4$
atoms from a BEC with no discernable uncondensed fraction in a
magnetic trap with $\omega_{\perp}/2\pi = 24$\,Hz and
$\omega_{\parallel}/2\pi = 8$\,Hz and Thomas Fermi (TF) radii of $
\approx 13\,\mu$m and $ \approx 32\,\mu$m, respectively. The
vertical lattice was slowly raised in 250\,ms, forming an array of
2D ``pancakes." After the first 100\,ms of this loading, we raised
the $\lambda$-lattice in the remaining 150\,ms \footnote{We load to
$\simeq 12\,E_R$ in 50\,ms then to $\simeq 100\,E_R$ in 100\,ms.}.
%
Ideally, at zero temperature, sufficiently slow loading produces a
vertical array of 2D Mott insulating systems with one atom in
$\left|g\right>$ of each $\lambda$-site. To prepare Poisson states
with $\left<N\right> \approx 1$ in $\left|g\right>$, we used the
same procedure, except we loaded both lattices quickly
(500\,$\mu$s). This ``fast" load is adiabatic only compared to band
excitation \cite{SebbyStrabley2006}. We destroyed the phase
coherence between atoms in different $\lambda$-sites by holding
atoms in the lattice for 5 ms. After preparing either of these
initial states, we applied the NI-BS by transforming to the
$\lambda/2$-lattice in 300 $\mu$s, a timescale fast enough to
maintain coherence within a double well, yet slow enough to avoid
vibrational excitation. After a variable hold time $t$, the atoms
were released and imaged after 13 ms TOF.

\begin{figure}[ptb]
\includegraphics[scale=.50]{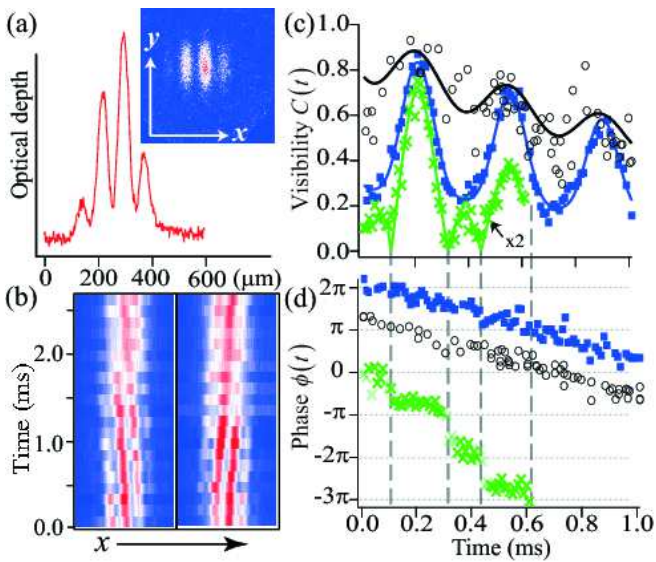}\caption{(color online) (a) A double-slit interference pattern (inset) and
integrated density profile after 13\,ms TOF.  (b) Example
$y$-integrated diffraction pattern as a function of time for tilts
$V=4.90(3)\,$kHz (left) and $V=-4.90(3)\,$kHz (right). (c) $C(t)$
for $N=1$ (black $\circ$), $N=2$ (green $\times$), and
$\left<N\right>=1$ (blue $\blacksquare$). Solid lines are a fit to
Eq.~\ref{ct}. Note: The plotted $C(t)$ for $N=2$ was scaled by 2.
(d) $\phi(t)$ for the cases shown in (c). The $\phi(t)$ points which
were derived from a signal with $2\times C(t) < 0.1$ are shaded in
light green.
$\phi(t=0)=0$ for all three cases, but each has been shifted for visual clarity. }%
\label{dslit}%
\end{figure}

Fig.~\ref{dslit}a shows an example double-slit interference pattern.
We extract the fringe visibility $C(t)$ and the spatial phase
$\phi(t)$ by fitting the profile to:
\begin{equation}
F(x,t)=Ae^{-\frac{(x-x_{0})^2}{2\sigma^{2}}}
    \left[ 1+ C\left(
t\right) \cos\left( \frac{x-x_{0}}{\Delta} + \phi(t)\right)
\right],\nonumber
\end{equation}
where $\sigma$ is determined by the width of the ground state
wavefunction in a $\lambda$-site, and $2 \pi \Delta$ is the fringe
spacing. We use $C(t)$ to measure the number statistics of our input
states. $\phi(t)$ evolves linearly due to an energy offset (tilt)
$V$ between the sites within a double well
\footnote{ $V$ can result from misalignment of the
$\lambda/2$-lattice~\cite{SebbyStrabley2006}, which we can in
principle compensate with a small amount of the $\lambda$-lattice.
The residual $\approx 1\,$kHz tilt in Fig.~\ref{dslit}d had no
effect on the results presented here.} (see Fig.~\ref{dslit}b).
%
%
%

The NI-beam splitter binomially splits $N$ atoms in each
$\lambda$-lattice site into superpositions of states
$\left|n_L,n_R\right>$, with $n_L$ atoms in $\left|L\right>$ and
$n_R=N-n_L$ atoms in $\left|R\right>$, resulting in the time
dependent wavefunction
\begin{equation}\label{wf}
\left|\psi(t)\right> = \sum_{n_L=0}^N c(n_L,
n_R)\left|n_L,n_R\right> e^{-i \omega(n_L,n_R) t}
\end{equation}
where $\left|c(n_L,n_R)\right|^2 = \left(\frac{1}{2}\right)^N
\left(N!/n_L!n_R!\right)$. In the two-mode approximation
interactions and tilt determine $\hbar \omega(n_L,n_R) = n_L V +
(1/2) U n_L(n_L-1) + (1/2) U n_R(n_R-1)$, where $U$ is the on-site
interaction energy per particle.  For our system $U/h \approx 3$
kHz.

In the absence of interactions and inhomogeneities,
$\left|\psi\right>$ always factors into products of single particle
states, $\left|\psi\right> =
\left(\left|1,0\right>+e^{i\omega_{\rm{tilt}}t}\left|0,1\right>\right)^N$,
where $\omega_{\rm{tilt}} = V/\hbar$. All the terms in Eq.~\ref{wf}
constructively interfere and $C(t)$ is maximized and constant. When
interactions are included, $\left|\psi\right>$ is not always
factorable, and $C(t)$ (for atoms in the ground band) is modulated
by $\left|\cos^{N-1}( U t/\hbar)\right|$ \cite{Johnson2006,WALLS1997}. Summed
over a distribution of site occupation numbers, the total $C(t)$ is
given by
\begin{equation}\label{ct}
C(t) = \frac{e^{-\Gamma t}}{\sum_{N=1}^{N_{\rm{max}}} f_N
N}\left|\sum_{N=1}^{N_{\rm{max}}} f_N N \cos^{N-1}(U t/\hbar)\right|
\end{equation}
where $f_N$ is the fraction of $\lambda$-sites with $N$ atoms (for
our case we limited $N_{\rm{max}}=4$), $\sum f_N = 1$, and $\Gamma$
is
an empirical dephasing rate that
we attribute to lattice inhomogeneities.
Fitting $C(t)$ with Eq. \ref{ct}, we can extract the relative
fractions $f_N$.

\begin{figure}[ptb]
\includegraphics[scale=.50]{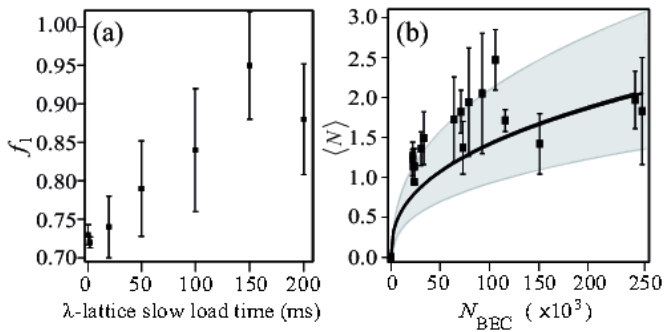}\caption{We use Eq.~\ref{ct} to fit
$C(t)$ for the fast and slow load data. (a) Extracted $f_1$ for
varied slow load times.  The uncertainties are from the fit to Eq.
\ref{ct}. (b) Using Eq.~\ref{ct} to fit the fast load $C(t)$ we
extract values for $f_1,f_2,f_3$ and $f_4$.  From those coefficients
we determine $\left<N\right>$, the peak number density of atoms in
$\lambda$-lattice sites, as a function of $N_{\mathrm{BEC}}$. The
line and the grey shaded area are the peak number density scalings
predicted by the TF-approximation (with no adjustable parameters)
and our estimated uncertainty in the calibration of
$N_{\mathrm{BEC}}$ (factor of $\pm 1.5$). The uncertainties on the
individual points are from a combination of the uncertainties
in $f_i$ and from the fit determining $\left<N\right>$.}%
\label{mott}%
\end{figure}

For the slow loading case with $N_{\rm{BEC}}\leq 2\times 10^4$
(average filling $\leq 1$ per site) we expect to have a Mott state;
we measure $C(t)$ to be approximately constant (see Fig.
\ref{dslit}c), indicating that occupied $\lambda$-lattice sites have
$N=1$. ($C(t)$ shows a slow decay, which we attribute to tilt
inhomogeneities.) Extracting $f_N$ from $C(t)$, we determine
$f_1=0.94(6)$. (As shown in Fig.\,\ref{mott}a, $f_1$ was optimized
at a load time of 150\,ms.) The uncertainty in $f_1$ is dominated by
the shot-to-shot scatter in the double-slit visibility $C$, which is
significantly larger than for fast loading. We do not understand
this increased noise, but note that the shot-to-shot scatter in
$\phi$ (Fig.\,\ref{dslit}d) is not increased. This suggests an
increased sensitivity to an uncontrolled initial condition, e.g.,
number or temperature.

The near-zero value of $f_2$ indicates a strongly number-squeezed
initial state, as expected in a Mott insulator (compared to the
Poisson value of $0.3$ for average filling of unity, ignoring
$f_0$). The value of $f_1$ does not, however, necessarily represent
the fidelity of the Mott state since this measurement is insensitive
to sites with $N=0$. Assuming $f_1=0.94$ and $f_2=0.06$ and a
homogeneous system, we can roughly estimate the temperature $T$ in
the $\lambda$-lattice from $f_2=e^{-U/k_B T}$. This gives $T \leq
0.36 U/k_B\simeq50$\,nK.

For the fast loading case where we expect a Poisson distribution of
site occupation numbers, there can be $N>1$ atoms at any
$\lambda$-site. In this case we find that $C(t)$ shows distinct
collapses and revivals,
caused by interactions (see Fig.~\ref{dslit}c). Only at integer
multiples $l$ of $T_{\mathrm{rev}}=h/U$ does the wavefunction within
a site factor: $\psi = \left( \left|1,0\right>
+e^{il\omega_{\rm{tilt}}T_{\rm{rev}}} \left|0,1\right>\right)^N$. We
observe revivals in $C(t)$ with a period of $\approx 350\,\mu$s, in
good agreement with our calculated value of $U/h = 2.88$\,kHz. While
similar to the interaction-induced collapse and revival seen in
\cite{Greiner2002b}, the interference here is between the sites of a
double well and is created by a topology change in the lattice,
rather than between sites extending over the entire lattice and
created by changing the lattice depth. This allows us to see
interference even in the absence of global coherence, as evidenced
by the interference pattern generated from the highly squeezed $N=1$
slow load state.

Using Eq.~\ref{ct} to fit the fast load $C(t)$, we extract the
number distribution of atoms in $\lambda$-lattice sites. By fitting
that distribution to an average over Poisson distributions whose
local mean occupation numbers are proportional to the initial
TF-density profile, we extract $\left<N\right>$, the peak number
density of atoms in $\lambda$-lattice sites.
This weighted Poissonian distribution fits the data within
experimental uncertainty.
Fig. \ref{mott}b shows $\left<N\right>$ as a function of
$N_{\mathrm{BEC}}$
%
%
and the power law scaling ($N_{\rm{BEC}}^{2/5}$) for
$\left<N\right>$ expected from the TF-approximation.
%
%
For large values of $N_{\mathrm{BEC}}$ the data deviates from the
trend set by the lower values of $N_{\mathrm{BEC}}$. While not
completely understood, this result may be attributed to three-body
loss mechanisms and a larger thermal fraction for larger values of
$N_{\mathrm{BEC}}$.


We now consider input states with $N=2$ atoms in $\left|g\right>$ of
the $\lambda$-lattice sites. If we simply loaded the
$\lambda$-lattice from our harmonic trap with increased initial peak
density, the resulting ``shell structure" should produce a
significant number of sites with $N=1$ surrounding the central core
of $N=2$ \cite{Folling2006, Campbell2006}, limiting the maximum
value of $f_2$. In principle, the flexibility of the double well
lattice can be used to construct a purer $N=2$ state in the
$\lambda$-lattice by combining two $\lambda/2$-lattice sites:  we
first load the vertical lattice as in the slow load case above, but
with an initial number ($N_{\mathrm{BEC}} \approx 7\times 10^4$)
large enough to produce $\simeq 1$ atom per site in the
$\lambda/2$-lattice. We then raise the $\lambda/2$-lattice in
100\,ms (ideally, slow enough to make a Mott insulating state with
one atom in the ground state of each $\lambda/2$-site). Finally we
adiabatically (in $30$\,ms) combine pairs of $\lambda/2$-sites into
$\lambda$-sites, thus performing an I-BS. The input states are twin
Fock states of one atom each in $\left|L\right>$ and
$\left|R\right>$,
the two-atom interacting ground state of the $\lambda/2$-lattice.
The output modes are $\left|g\right>$ and $\left|e\right>$ states of
$\lambda$-sites. Interactions and adiabaticity
ensure that both atoms go into $\left|g\right>$, the two-atom ground
state of the $\lambda$-lattice.
This ``constructed pair" state is then input to our interferometer
(split with an NI-BS).

As shown in Fig.~\ref{dslit}c,d we find that after using the
constructed pair technique to load the $\lambda$-lattice, $C(t)$ and
$\phi(t)$ show unique features which can only be attributed to
$\lambda$-sites with $N=2$. As in the fast load case, $C(t)$
collapses and revives, but there are revivals at half the period
compared to the fast load case ($T_{\mathrm{rev}}/2$ rather than
$T_{\mathrm{rev}}$). Like the slow and fast load cases, $\phi(t)$
evolves linearly as $\omega_{\rm{tilt}}$ (here $\approx 0$), but
here $\phi(t)$ makes clear jumps between revivals. These jumps are
$\simeq\pi$, particularly if only the $\phi(t)$ points with $2
\times C(t)
> 0.1$ are considered.  At $lT_{\mathrm{rev}}$ the wavefunction can
be factored as $\left|\psi\right>=\left(\left|1,0\right> +
e^{i\omega_{\rm{tilt}}T_{\mathrm{rev}}} \left|0,1\right>\right)^2$,
so $C(t)$ revives.  Unlike the fast load case, at
$lT_{\mathrm{rev}}/2$, $C(t)$ also revives but with the wavefunction
factored into oppositely-phased single particle states,
 $\left|\psi\right> =\left(\left|1,0\right> -
e^{i\omega_{\rm{tilt}}T_{\mathrm{rev}}/2}\left|0,1\right>\right)^2$,
giving a $\pi$ phase-shifted interference pattern~\cite{Johnson2006,WALLS1997}.

Our maximum measured $C$ in Fig.~\ref{dslit}c for the constructed
pairs technique is a factor of $\approx$ 2 less than the fast and
slow load cases, and our measured $C(T_{\mathrm{rev}}/2)$ is lower
than $C(T_{\mathrm{rev}})$.
These differences
can be largely explained
%
%
by the fact that after the I-BS, a large fraction of atoms ($\simeq$
30$\%$) are in $\left|e\right>$ (revealed by imaging the
$\lambda$-lattice BZ),
and make an interference pattern $\pi$-out-of-phase with that made
by atoms in $\left|g\right>$
\footnote{
%
%
Eq.~\ref{ct} does not account for $\left|e\right>$ population.
}.
%
%
%
After the NI-BS the incoherent sum
of sites with atoms equally in $\left|g\right>$ and
$\left|e\right>$---in total $\simeq$ 60$\%$ of the atoms---thus
contributes $C(t)=0$.
$\left|e\right>$-population could be caused by I-BS input states
such as $\left|n_L,n_R\right> = \left|2,0\right>$,
$\left|2,1\right>$, and $\left|1,0\right>$, all of which would
reflect impurities in the initial $\lambda/2$-lattice Mott state.
The presence of atoms in $\left|e\right>$, however,
%
%
also points to a unique capability of the double well lattice:
traditionally it has been impossible to detect empty sites, but by
using the constructed pairs technique to load the $\lambda$-lattice
and measuring $N=1$, we can detect the presence of empty sites in
the initial $\lambda/2$-lattice.

Using the versatility and control of the double well lattice,
future studies can also be focused on creating and exploring more
complicated input states with interferometry, such as entangled
$\left|N,0\right>+\left|0,N\right>$ states. Using the constructed
pairs technique but with a NI-BS to combine pairs of
$\lambda/2$-sites to single $\lambda$-sites, we can create
$\left|2,0\right>+\left|0,2\right>$ in the $\left|g,e\right>$
basis. This would be the atom optics equivalent of the
Hong-Ou-Mandel effect \cite{hom}.

\begin{acknowledgments}
This work was supported by DTO, ONR, and NASA. The authors thank
Jens Kruse, Ian Spielman, and Steve Rolston for contributions to the
project. JS-S, BLB, and PJL acknowledge support from the NRC.  PRJ
acknowledges support from the IC Postdoctoral Program.
\end{acknowledgments}

\newcommand{\noopsort}[1]{} \newcommand{\printfirst}[2]{#1}
   \newcommand{\singleletter}[1]{#1} \newcommand{\switchargs}[2]{#2#1}

\end{document}